# Content-based Fine-grained Flow Management Supporting Out-of-Path Transparent Add-ons

Anna Ishizaki
*Graduate School of Science and Technology,*
*Keio University*
3-14-1 Hiyoshi, Kohoku, Yokohama, Kanagawa 223-8522, Japan
anna@west.sd.keio.ac.jp

Takuma Fukui
*Graduate School of Science and Technology,*
*Keio University*
3-14-1 Hiyoshi, Kohoku, Yokohama, Kanagawa 223-8522, Japan
fukui@west.sd.keio.ac.jp

Hiroaki Nishi
*Department of System Design,*
*Faculty of Science and Technology,*
*Keio University*
3-14-1 Hiyoshi, Kohoku, Yokohama, Kanagawa 223-8522, Japan
west@sd.keio.ac.jp

*Abstract*—This study aims to realize a mechanism for packet processing in the edge domain while maintaining network transparency, in order to accommodate diverse service requirements in smart communities. Since conventional flow control, which operates on a per-packet basis, lacks flexibility, we propose a content-based fine-grained flow management method that enables control at the level of individual content segments within packets. In addition, we introduce an out-of-path transparent add-on architecture to address the limitations of conventional transparent add-ons, which assume the presence of processing resources on the main path. The proposed system implements one approach for selective content masking and two approaches for out-of-path anonymization. Furthermore, we develop a mechanism for dynamically rewriting Ack and Seq numbers to preserve TCP session integrity. The proposed approaches were implemented and evaluated on Mininet, and the results demonstrate that effective flow management can be achieved with minimal impact on network delay while maintaining network transparency.

## I. Introduction

This study focuses on the concept of smart communities, which enhance, sophisticate, and integrate multiple societal infrastructures through the application of Information and Communication Technology (ICT). Traditionally, information within individual infrastructures has been handled separately. However, in smart communities, the cross-utilization of information across various infrastructures is being promoted, resulting in more diverse and complex services. To support such services effectively, it is essential to place each service at an appropriate processing location. A smart community consists of the IoT device domain, the edge domain, and the cloud. The edge domain refers to the intermediate area through which data passes from IoT devices to the cloud.

Edge computing enables low-latency data processing in the edge domain, allowing privacy-preserving tasks such as data anonymization before sending data to the cloud, thereby enabling the realization of hierarchical privacy protection. It is useful for services using personal data, as it protects privacy before reaching the cloud. In addition, research on cryptographic techniques and key management for concealing data from the cloud has also been conducted [1], and edge computing plays a vital role as a foundational technology for privacy protection in smart communities.

On the other hand, Internet of Things (IoT) devices are generally resource-constrained and exist in large numbers, making large-scale updates impractical. Therefore, network-based processing must be designed in such a way that it does not affect the operation of IoT devices. In this context, network transparency, a property that conceals the existence of network-side services from users, emerges as a critical requirement. To meet this requirement, the concept of transparent add-ons has been proposed. Transparent add-ons provide services by augmenting functionality at the edge domain without requiring any modifications to IoT devices. To realize such transparent add-ons, a packet stream analysis method called Authorized Stream Contents Analysis (ASCA) [2] has been proposed. ASCA targets not only the packet headers but also the payload, enabling service provisioning based on the actual content of the packets.

In addition, flow management in the edge domain is a crucial aspect. A flow table, typically used in such management, is a routing control table that distinguishes packet streams based on five tuples, which consist of the source IP address, destination IP address, source port, destination port, and protocol type. It specifies actions such as forwarding or dropping packets for each identified stream. Shimahara et al. [3] proposed an approach to flow control, in which not only the 5-tuple of packet headers but also application-layer content is used as a flow identifier. Based on this information, their system controls the forwarding or dropping of packets.

Based on the discussion above, this study addresses the three key challenges.

・First, conventional flow management provides only coarse-grained control at the content level, making it difficult to manage data on a per-user basis, even when multiple users' data are contained within a single packet. This limitation becomes particularly problematic when users have different data-handling contracts, as it requires distinct processing for each user's data. However, such per-user differentiation is not feasible under packet-level control, which lacks the granularity to isolate and manage individual content units.

・Second, conventional transparent add-ons assume that processing resources are available on the on-path nodes, which refer to nodes located along the communication path from the IoT device to the cloud. In practice, however, such on-path nodes may lack sufficient computational capacity. To address this limitation, this study introduces a flexible mechanism for path branching, enabling offloading of processing to out-of-path nodes.

・Third, to enable content-based processing, it is often necessary to rewrite the packet payload. Moreover, in scenarios involving out-of-path transparent add-ons, appropriate path branching must be performed to forward specific data portions to external processing nodes. However, such operations, including payload rewriting and path branching, can interfere with Transmission Control Protocol (TCP) functionalities, particularly sequence and retransmission control, potentially disrupting the maintenance of active TCP connections.

To address these issues, this study outlines four objectives:

(1) Implementation of Network-Transparent, Fine-Grained Content-Based Flow Management.
(2) To realize out-of-path transparent add-ons, by enabling dynamic path branching based on content attributes.
(3) To execute content-level processing while maintaining active TCP connections, through dynamic rewriting of Ack and Seq numbers and the establishment of new TCP connections when necessary.
(4) To minimize the impact on network performance, taking into account the possibility of increased transmission delay resulting from the above mechanisms.

Specifically, the network is designed such that the link delay between nodes in the smart community is assumed to be 5 ms, and the one-way delay does not exceed 150 ms, in accordance with the quality requirements defined in the ITU-T Recommendation G.114 [4], which states that delays under 150 ms are acceptable for most applications.

## II. RELATED STUDIES

### A. Software Defined Network (SDN)

Software-Defined Networking (SDN) is a network architecture that decouples the control and data planes, enabling centralized and flexible network management through software. OpenFlow is a representative implementation of SDN, where controllers perform control logic and switches handle packet forwarding based on flow table rules. Leveraging SDN's flexible control capabilities, various applications such as dynamic load balancing [5] have been proposed.

However, a limitation of conventional SDN-based approaches is that they often assume that all participating devices, including end devices, are SDN-compatible. This requirement contradicts the principle of network transparency, particularly for resource-constrained devices such as IoT nodes. In contrast, the system proposed in this study does not require end devices to be SDN-compatible, thus maintaining network transparency while enabling fine-grained content-based control within the network.

In addition, Intel's Tofino [6], a high-performance programmable switch using P4, offers flexible packet processing. However, it does not support application-layer payload analysis or modification, which is essential for the content-based processing addressed in this study.

### B. Packet Rewriting

As a technique used to realize a transparent add-on, a method for rewriting packet contents has been proposed [7]. In this approach, an intermediate node between the IoT device transmitting the data and the cloud first analyzes the packet header upon packet arrival. If a packet is identified as originating from an IoT device and using TCP, payload analysis is performed. When the payload contains the string "POST" and power consumption data, the packet proceeds to a rewriting process in which Differential Privacy (DP) is applied. Laplace noise is added to the numerical data, and after

Fig. 1. Content table example

recalculating the checksum, the modified packet is forwarded to the cloud.

However, this process alone cannot handle changes in the payload size. Therefore, this study additionally supports modifications to the payload size.

## III. PROPOSED SYSTEM

This study proposes a framework for content-based flow management in smart communities, in which packets are processed on a per-content basis.

### A. Assumed Environment

This study assumes a smart community network composed of three domains: the IoT device domain, the edge domain, and cloud. The IoT device domain includes devices such as sensors and smartphones that transmit potentially sensitive data to the cloud. In the cloud, servers manage both data storage and user contract information. The edge domain hosts intermediate nodes that transparently process packets using content tables, which are updated based on user contracts.

Intermediate nodes in the edge domain analyze the packet payload in a transparent manner and perform content-level processing as necessary, based on the content table. This study proposes two types of content processing. The first is content masking, in which only the specified content is removed from a packet even when multiple contents are present, ensuring that non-targeted content remains unaffected. The second is out-of-path anonymization. Implementing anonymization functionality uniformly across all intermediate nodes is inefficient due to constraints in computational resources and energy consumption. Therefore, this study assumes that anonymization functions are not installed in on-path nodes and proposes a method in which the data is branched out-of-path, processed by a dedicated out-of-path node, and then forwarded to the cloud.

To realize these content processing methods, this study proposes one approach for content masking and two approaches for out-of-path anonymization. For the out-of-path anonymization process, it is assumed that the cloud specifies in advance which approach should be applied, depending on the situation.

### B. Content Table

To realize content-based flow management, a finer-grained content table than the conventional flow tables mentioned above is introduced. An example of transmitted data and its corresponding content table is shown in Fig. 1. Since the processing decisions are made solely based on the content of the payload, there is no need to specify the 5-tuple. For content specification, it is required to precisely identify the relevant portion within the packet's payload. However, once this condition is satisfied, service providers can freely define the content, thereby enabling flexible and adaptable flow management.

## C. Content Masking

This study proposes the Shortening Approach shown in Fig. 2 as a method for selective content masking. In this approach, unnecessary content is removed from the packet payload, and the packet is forwarded to the next node after recalculating the Total Length field and checksums in the Internet Protocol (IP) and TCP headers. However, in TCP communication, when the payload is reduced at an intermediate node, the Ack number returned by the server becomes smaller than what the IoT device expects. Consequently, the IoT device may retransmit data that has already been masked, potentially leading to a loop in which the same data is repeatedly retransmitted and masked. Such a loop not only wastes network bandwidth but may also degrade the performance of other data transmissions. Therefore, it is essential that the intermediate node accurately tracks the reduction in data length and appropriately rewrites the Ack and Seq numbers in the TCP header to maintain correct TCP session behavior. This mechanism is also applicable when the data size increases, such as in content modification scenarios, by similarly adjusting the length field and both the Seq and Ack numbers to reflect the updated payload size.

Alternatively, there exists an approach that retains the original payload length by applying zero-padding to the masked portions. This method simplifies processing, as it requires only the recalculation of checksums. However, because the unnecessary content remains part of the packet size, this approach leads to inefficient use of network bandwidth. Considering these trade-offs, this study adopts the Shortening Approach to achieve efficient and content-aware flow control.

## D. Out-of-path Anonymization

This study proposes two methods for realizing out-of-path anonymization. In the first method, the "*Integrated Approach*" shown in Fig. 3, when a packet contains data that requires special processing, only the relevant portion is sent to an out-of-path dedicated processing node. Once the processing is completed, the result is returned, integrated with the remaining data, and forwarded to the next node. This approach has the advantage of preserving privacy by transmitting only the relevant data while maintaining the original packet flow. On the other hand, a limitation of this approach is that data not subject to dedicated processing must temporarily wait until the processing is completed, which can reduce processing efficiency.

In contrast, the second method, the "*Non-Integrated Approach*" shown in Fig. 4, also sends only the relevant portion of the data to the dedicated processing node, but the processed data and the remaining data are not reintegrated. Instead, each is independently transmitted to the cloud. This approach is more efficient, as the data that does not require processing can be sent to the cloud without waiting. However, since the processed data is transmitted via a separate TCP connection between the dedicated processing node and the cloud, the source information in the packet header (such as IP address and port number) reflects the processing node rather than the original IoT device. Therefore, the use of this approach assumes that the cloud can identify the origin of the data based on the content in the payload. Additionally, this approach provides an advantage that is not available in the Integrated Approach: the cloud can receive data sequentially as soon as edge-side processing is completed.

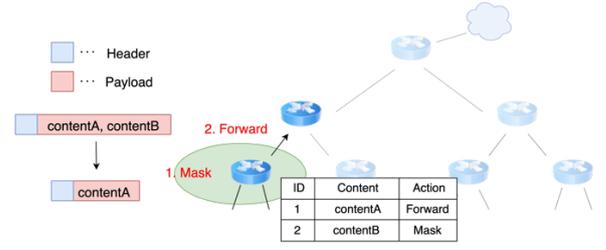

Fig. 2. Shortening Approach

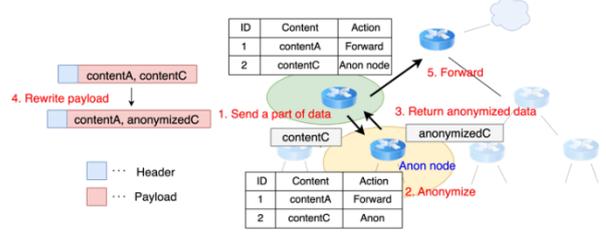

Fig. 3. Integrated Approach

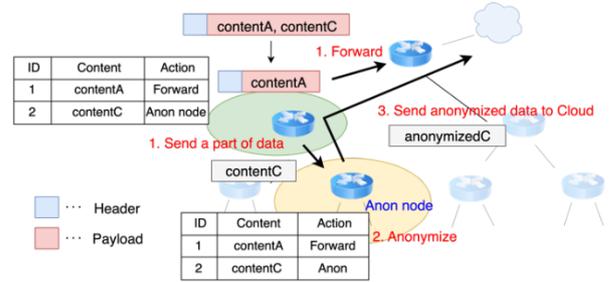

Fig. 4. Non-Integrated Approach

The Integrated Approach avoids extra overhead but requires waiting for reintegration. In contrast, the Non-Integrated Approach reduces delay by sending data separately, at the cost of increased overhead due to the increased number of packets.

## IV. IMPLEMENTATION

We used Mininet [8], to build the environment shown in Fig. 5, implementing each node in C++. TCP communication is used between the IoT device and the cloud. The IoT device sends power data as shown in Fig. 1, and the cloud receives data from multiple nodes. Users define which fields to mask or anonymize at registration, but the IoT device sends all data as-is, maintaining network transparency. The specifications of the physical machine used in the experiments are shown in TABLE I.

The processing flows for each approach implemented in the edge domain are illustrated in Algorithms 1–6. In the algorithms, Δx represents a value managed internally by each intermediate node for each TCP connection, which accumulates the total size of packets that have been shortened. This value is maintained until the corresponding TCP connection is terminated.

Upon packet arrival, the header is analyzed to determine whether the packet belongs to a TCP connection established between the IoT device domain and the cloud. If confirmed, the intermediate node rewrites the Ack and Seq numbers as needed, enabling the TCP connection to remain consistent even if the payload size is altered during processing. When a packet is being sent from the IoT device domain to the cloud, the intermediate node references the content table to determine whether the packet contains any content that requires processing. This initial procedure is common to all approaches. However, the subsequent steps vary depending on the selected approach. As flow table distribution is addressed in [3], we assume the content table is preconfigured by a controller.

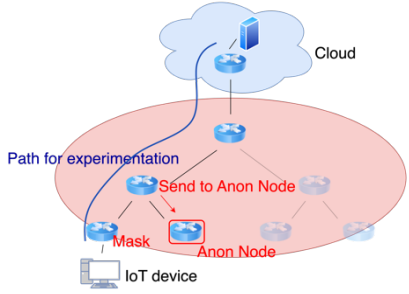

Fig. 5. Assumed environment

**Algorithm 1:** Common to three approaches
**Data:** ReceivedPacket $p$
1  $\Delta x \leftarrow 0$ // Total bytes shortened
2  **if** $p.protocol \neq TCP$ **or** $\neg$ BETWEEN_CLOUD_AND_IOT($p$) **then**
3  | FORWARD($p$);
4  | **return**;
5  **if** $p.direction = IoT \to Cloud$ **then**
6  | **if** $\Delta x \neq 0$ **then**
7  | | $p.\text{seq} \leftarrow p.\text{seq} - \Delta x$;
8  | | UPDATE_CHECKSUMS($p$);
9  | **if** $p.flags = PSH$ and $ACK$ **then**
10 | | REFER_TO_CONTENT_TABLE($p$);
11 | | SELECTED_APPROACH($p$); // $\to$ Alg. 2, 3, or 5
12 **else**
13 | **if** $\Delta x \neq 0$ **then**
14 | | $p.\text{ack} \leftarrow p.\text{ack} + \Delta x$;
15 | | UPDATE_CHECKSUMS($p$);
16 FORWARD($p$);

**Algorithm 2:** Shortening Approach
1  **if** HAS_MASK_TARGET($p$.payload) **then**
2  | $bytesRemoved \leftarrow$ DELETE_AND_SHIFT($p$.payload);
3  | $\Delta x \leftarrow \Delta x + bytesRemoved$;
4  | UPDATE_TOTAL_LENGTH_AND_CHECKSUMS($p$);

**Algorithm 3:** Integrated Approach
1  **if** HAS_OUTNODE_DATA($p$.payload) **then**
2  | ENSURE_CONNECTION($outNode$);
3  | $partial \leftarrow$ EXTRACT_OUTNODE_DATA($p$.payload);
4  | SEND($outNode, partial$) // $\to$ Alg. 4: Send and wait
5  | $updated \leftarrow$ WAIT_RESPONSE($outNode$);
6  | $p.\text{payload} \leftarrow$ MERGE($updated$, REMAINING($p$));
7  | $\Delta x \leftarrow \Delta x + (|partial| - |updated|)$;
8  | UPDATE_TOTAL_LENGTH_AND_CHECKSUMS($p$);

**Algorithm 4:** OutNode Process in Integrated Approach
**Data:** ReceivedData $d$
1  REFER_TO_CONTENT_TABLE($d$);
2  **if** HAS_ANONYMIZABLE_CONTENT($d$) **then**
3  | ANONYMIZE($d$);
4  **return** $d$ // Return to original sender

In the Shortening Approach (Alg. 1, 2), the designated content is removed from the payload. After updating X, the intermediate node recalculates the checksums in both the IP and TCP headers and updates the Total Length field in the IP header before forwarding the packet to the next node.

In the Integrated Approach (Alg. 1, 3, 4), a new TCP connection is established between the on-path node and the out-of-path node, and only the target content is sent to the out-of-path node. The on-path node waits until it receives a response from the out-of-path node indicating that anonymization has been completed. The anonymized content is then integrated with the remaining unsent data. If the payload size is changed due to anonymization, the on-path node performs the same processing as in the Shortening Approach (i.e., updating X and rewriting the headers) before forwarding the packet to the next node.

TABLE I. DEVICE CONFIGURATION DETAILS

| Device | Shuttle DS81D |
|---|---|
| OS | Ubuntu 22.04.5 LTS |
| CPU | Intel(R) Core(TM) i5-4590S CPU @ 3.00GH |
| Memory | 7.6 GiB |

**Algorithm 5:** Non-Integrated Approach
1  **if** HAS_OUTNODE_DATA($p$.payload) **then**
2  | ENSURE_CONNECTION($outNode$);
3  | $partial \leftarrow$ EXTRACT_OUTNODE_DATA($p$.payload);
4  | SEND($outNode, partial$) // $\to$ Alg. 6: Send and don't wait
5  | $p.\text{payload} \leftarrow$ REMOVE($partial$, $p$.payload);
6  | $\Delta x \leftarrow \Delta x + |partial|$;
7  | UPDATE_TOTAL_LENGTH_AND_CHECKSUMS($p$);

**Algorithm 6:** OutNode Process in Non-Integrated Approach
**Data:** ReceivedData $d$
1  REFER_TO_CONTENT_TABLE($d$);
2  **if** HAS_ANONYMIZABLE_CONTENT($d$) **then**
3  | ANONYMIZE($d$);
4  ENSURE_CONNECTION($cloud$);
5  SEND($cloud, d$);

In contrast, in the Non-Integrated Approach (Alg. 1, 5, 6), only the content requiring anonymization is sent to the out-of-path node, while the remaining unprocessed data is immediately forwarded to the next node, thus completing the packet processing at the on-path node. The out-of-path node performs the anonymization and establishes a new TCP connection with the cloud to send the processed data.

In all approaches, the intermediate node first extracts the TCP payload by parsing the Ethernet, IP, and TCP headers of each packet. Once the payload is identified, the system uses regular expressions to locate the position of the target content within the payload for masking or anonymization. This allows precise selection and modification of data, while preserving the surrounding content. The anonymization process performed at the out-of-path node adopts the Laplace mechanism of DP, as previously employed in the implementation by Sato et al [7].

## V. EVALUATION AND RESULTS

### A. Processing Time

For each of the proposed methods, the processing time was measured. A link delay of 5 ms was configured for each link, and the average was calculated over 200 measurements. The arrival time of ACK packets was recorded using Wireshark [9]. The measurement intervals for the Shortening Approach, Integrated Approach, and Non-Integrated Approach are illustrated in Fig. 6, and the results of the measured processing times are presented in TABLE II.

In the measurement of the Shortening Approach, a 134-byte power data packet containing the content to be masked was used. According to the results shown in the TABLE II, the processing delay at the intermediate node accounted for approximately 8.4% of the transmission delay, indicating that the impact of intermediate node processing on overall transmission delay was minimal. The delay caused by rewriting the Ack number was about 0.11% of the ACK packet's transmission delay, suggesting that the effect of Ack number rewriting on latency is virtually negligible.

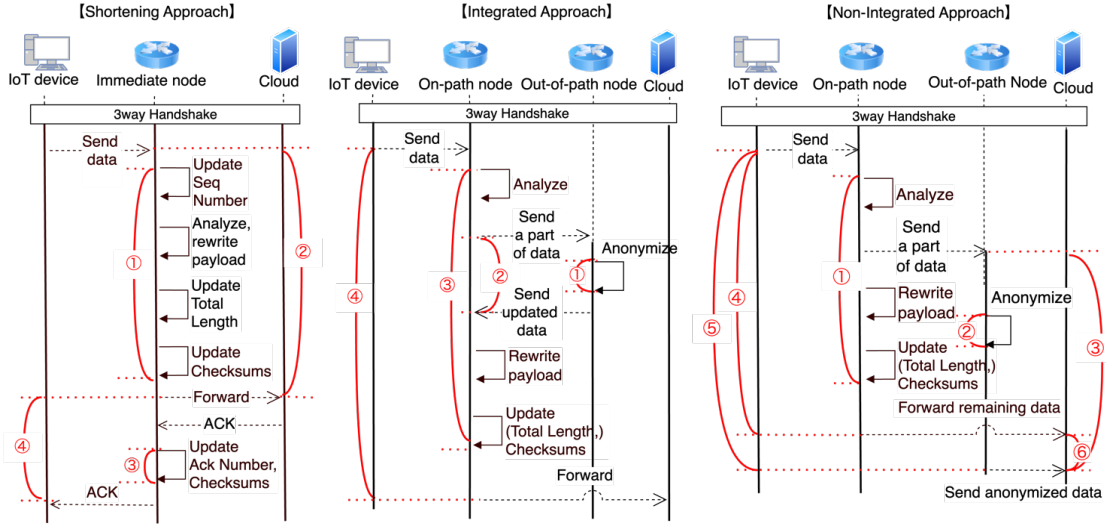

Fig. 6. The measurement intervals

In the evaluation of the out-of-path anonymization methods (Integrated and Non-Approaches), 130-byte power data containing content subject to out-of-path processing was used. The results show that the Integrated Approach incurs greater processing delay at the on-path node compared to the Shortening Approach, with approximately 78% of the delay caused by waiting for a response from the anonymization node. Since only about 15% of that time is spent on the anonymization process itself, the communication between the on-path and out-of-path nodes is the primary source of delay. In contrast, the Non-Integrated Approach enables non-anonymized data to be sent to the cloud approximately 13 ms earlier than anonymized data. Its transmission delay was comparable to that of the Shortening Approach, due to the elimination of waiting for a response from the anonymization node. Furthermore, the transmission delay of anonymized data remained equivalent to that of the Integrated Approach, indicating no disadvantage in separating transmissions. Overall, while the total time required for the cloud to receive all data is similar to other methods, the ability to deliver non-anonymized data earlier is a key advantage of the Non-Integrated Approach.

Furthermore, in all approaches, the transmission delay remained below 150 ms, confirming that the impact on the network was effectively minimized.

## B. Latency evaluation

We evaluated how increasing the data size from IoT devices affects transmission delay for each method. In the Shortening Approach, data size was increased from 134 to 1,072 bytes, along with a gradual increase in the size of content to be masked. The rate of delay increase was calculated based on the baseline at 134 bytes. For the Integrated and Non-Integrated Approaches, data ranged from 130 to 1,000 bytes, with delay increase calculated from the 130-byte baseline. In the Non-Integrated case, delays for anonymized and non-anonymized data were measured separately. All tests used a 5 ms link delay and 200 trials per data size. The results are shown in Fig. 7, Fig. 8. As a reference, we also measured delay without any flow management.

The highest observed delay increase was approximately 5.7% in the Integrated Approach, which remained sufficiently small relative to the increase in data size. This suggests that all proposed methods maintain stable transmission performance even as the data size increases. Furthermore, when comparing the Integrated and Non-Integrated Approaches, the Non-Integrated Approach showed more stable delay performance with respect to data size. In the Integrated Approach, the data is first split for processing, reassembled to its original size, and then synchronously transmitted to the cloud. This makes the transmission delay more sensitive to increases in data size. In contrast, the Non-Integrated Approach transmits each segment asynchronously without reassembly, limiting the impact of data size increases to each segment and thereby mitigating the overall transmission delay increase.

## C. Overhead of Path Branching

To evaluate the overhead introduced by path branching, we compared the transmission delays in two scenarios: one employing out-of-path anonymization, and the other performing on-path anonymization. In the on-path anonymization setting, anonymization was conducted directly at the on-path node without introducing path branching. We used 1,000-byte data packets and measured transmission delays under varying link delays between nodes. The results are presented in Fig. 9, while the average anonymization processing delays at the on-path and out-of-path nodes are summarized in TABLE III.

As shown in Fig. 9, the overhead caused by path branching was approximately 10 ms when the link delay was 1 ms, 15 ms with 5 ms link delay, and 27 ms with 10 ms link delay, across all approaches. Furthermore, according to TABLE III, the difference in processing delay between the on-path node and the out-of-path node was only about 0.4 ms, indicating that the impact of processing time difference on overall transmission delay is minimal. Therefore, the overhead introduced by path branching, when compared with the on-path configuration, is primarily attributed to additional link delay and the processing delay required for branching operations.

TABLE II. PROCESSING TIME OF THE THREE APPROACHES

| Section | Processing time [ms] | | |
| --- | --- | --- | --- |
| | Shortening | Integrated | Non-Integrated |
| 1 | 2.394 | 1.971 | 3.665 |
| 2 | 28.485 | 12.808 | 1.950 |
| 3 | 0.027 | 16.491 | 28.358 |
| 4 | 25.545 | 42.312 | 29.660 |
| 5 | - | - | 42.171 |
| 6 | - | - | 12.511 |

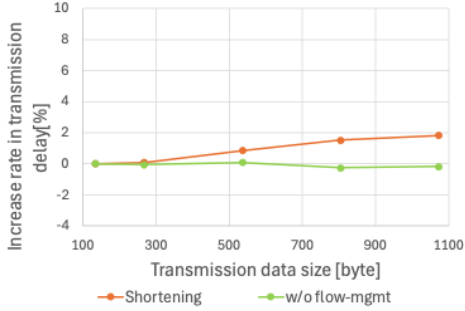

Fig. 7. Increase rate of transmission delay (content masking)

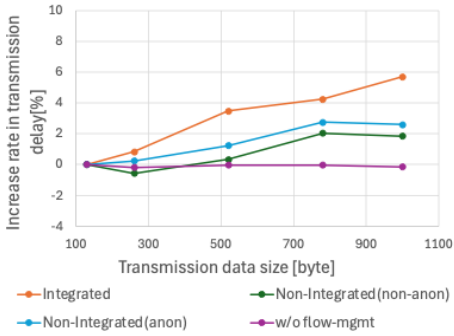

Fig. 8. Increase rate of transmission delay (out-of-path anonymization)

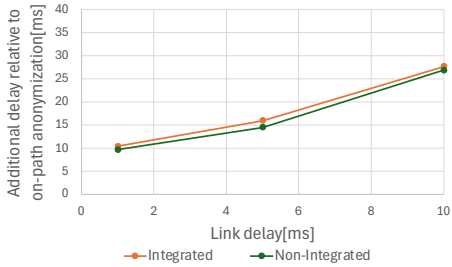

Fig. 9. Additional delay relative to on-path anonymization

TABLE III. ANONYMIZATION PROCESSING DELAYS

| Node | Anonymization processing time [ms] |
| --- | --- |
| On-path | 2.617 |
| Out-of-path | 2.219 |

In the implementation of the out-of-path transparent add-on in this study, the computational resources of the on-path and out-of-path nodes were configured to be equivalent, and the processing performed was not computationally intensive. However, in practical deployment environments, it is expected that high-load processing tasks will be offloaded to out-of-path nodes with greater computational capacity. In such scenarios, the processing delay difference between on-path and out-of-path nodes may become more significant, potentially resulting in greater overhead than observed in the experimental evaluation. Therefore, under such conditions, the path branching mechanism proposed in this study is effective. Moreover, since link delay has a substantial impact on the overhead, it is essential to make design decisions that take into account the state of the network links.

## VI. CONCLUSION

This study proposed a network-transparent, content-based flow management method for smart communities using packet rewriting. To handle cases where on-path resources are insufficient, we introduced out-of-path transparent add-ons that offload processing to external nodes. Two content processing methods were implemented, content masking and out-of-path anonymization. In content masking, specific payload content is selectively removed using a Shortening Approach while preserving TCP integrity by rewriting Ack and Seq numbers. Evaluation showed that content processing added only about 8.4% delay, with minimal impact on end-to-end performance. For anonymization, we designed a content-based method to avoid unnecessary data transmission to anonymization nodes. We proposed two strategies, Integrated and Non-Integrated Approaches. The Non-Integrated Approach improved efficiency by incrementally sending processed data to the cloud. Experiments confirmed that our system kept transmission delay under 150 ms, meeting latency requirements with minimal network impact.

ACKNOWLEDGMENT

These research results were obtained from the commissioned research (No. JPJ012368C08001) by National Institute of Information and Communications Technology (NICT), Japan.